\documentclass[%
 aip,
 amsmath,amssymb,
 reprint,%
]{revtex4-1}

\usepackage{graphicx}
\usepackage{dcolumn}
\usepackage{bm}

\usepackage[utf8]{inputenc}
\usepackage[T1]{fontenc}
\usepackage{mathptmx}
\usepackage{etoolbox}
\usepackage[dvipsnames]{xcolor}

\makeatletter
\def\@email#1#2{%
 \endgroup
 \patchcmd{\titleblock@produce}
  {\frontmatter@RRAPformat}
  {\frontmatter@RRAPformat{\produce@RRAP{*#1\href{mailto:#2}{#2}}}\frontmatter@RRAPformat}
  {}{}
}%
\makeatother
\begin{document}

\preprint{AIP/123-QED}

\title{Sub-nanosecond free carrier recombination in an indirectly excited quantum-well heterostructure}

\author{M. Perlangeli}
\affiliation{Dipartimento di Fisica, Università degli Studi di Trieste, 34127 Trieste, Italy}
\author{F. Proietto}
\affiliation{Dipartimento di Fisica, Università degli Studi di Trieste, 34127 Trieste, Italy}
\author{F. Parmigiani}
\affiliation{Dipartimento di Fisica, Università degli Studi di Trieste, 34127 Trieste, Italy}
\affiliation{Elettra-Sincrotrone Trieste S.C.p.A., 34149 Basovizza, Italy}
\affiliation{International Faculty, University of Cologne, Albertus-Magnus-Platz, 50923 Cologne, Germany}
\author{F. Cilento}
\email{federico.cilento@elettra.eu}
\affiliation{Elettra-Sincrotrone Trieste S.C.p.A., 34149 Basovizza, Italy}

\date{\today}

\begin{abstract}
Nanometer-thick quantum-well structures are quantum model systems offering a few discrete unoccupied energy states that can be impulsively filled and that relax back to equilibrium predominantly via spontaneous emission of light. Here we report on the response of an indirectly excited quantum-well heterostructure, probed by means of time and frequency resolved photoluminescence spectroscopy. This experiment provides access to the sub-nanosecond evolution of the free electron density, indirectly injected in the quantum-wells. In particular, the modelling of the time-dependent photoluminescence spectra unveils the time evolution of the temperature and of the chemical potentials for electrons and holes, from which the sub-nanosecond time-dependent electron density is determined. This information allows to prove that the recombination of excited carriers is mainly radiative and bimolecular at early delays after excitation, while, as the carrier density decreases, a monomolecular and non-radiative recombination channel becomes relevant. Access to the sub-nanosecond chronology of the mechanisms responsible for the relaxation of charge carriers provides a wealth of information for designing novel luminescent devices with engineered spectral and temporal behavior.
\end{abstract}

\maketitle
\section{\label{sec:sec1}Introduction}
Ultra-short light pulses can be used to transiently populate unoccupied states of matter and detect the relaxation dynamics by a suitable time-resolved spectroscopy. In such a scenario, Time-Resolved PhotoLuminescence (TR-PL) spectroscopy allows for the observation of the sub-nanosecond relaxation processes in the time domain \cite{McHale}, preserving the spectral information. This technique is particularly effective in atomic and molecular systems\cite{Bergman} and in solids with a direct band gap, where the radiative de-excitation pathway is favored\cite{Lakowicz, Cubeddu2002,Christensen2006,Piesch1990,Wagnieres1998}. In condensed matter, TR-PL has been applied predominantly to the study of semiconductors \cite{Klingshirn}, and, more recently, to the study of organic and inorganic perovskites \cite{Wieghold2020,Wieghold2020b,Franceschini2020}, and transition-metal dichalcogenides (TMDC) monolayers \cite{Manzeli2017,Tran2016,Brem2020,Li2020,Ruppert2014}. Interestingly, TR-PL can make available information complementary to those obtained by time-resolved optical spectroscopy. Their combination offers a comprehensive insight into both the radiative and non-radiative decay processes \cite{McHale}, the latter including elastic and inelastic electron-electron, electron-boson, and boson-boson scattering processes.
The PL dynamics from quantum-well states have been reported in previous studies, where different cases and materials have been discussed extensively. \cite{Nozik2001,LI20161809,doi:10.1063/1.3280384,Okamoto2005,Lefebvre1999}. 
\\
Here we report on a novel study of the TR-PL of a quantum-wells (QWs) heterostructure excited by ultrashort laser pulses with ~2.4 eV photon energy. The experiments have been made possible thanks to an ad-hoc experimental setup designed and built to provide simultaneous information in the time and frequency domains of the PL after the impulsive optical  excitation.
By analyzing the TR-PL spectra, we reveal the chronology of light emission from the QWs and barrier layers. Besides, by modelling the time evolution of the PL spectrum in terms of a Fermi-Dirac distribution for the photoexcited electron-hole plasma\cite{Modesti1980,MODESTI1981581,Modesti1981exp2}, the hot carrier recombination processes are unveiled. The structure of our sample allows to capture most of the electron-hole pairs excited by the photon pulse in essentially a single or in a few QWs in time shorter than the recombination time and therefore to study the system at high plasma densities.

\section{\label{sec:sec2}Methods and Results}
\begin{figure}
\includegraphics[width=\columnwidth]{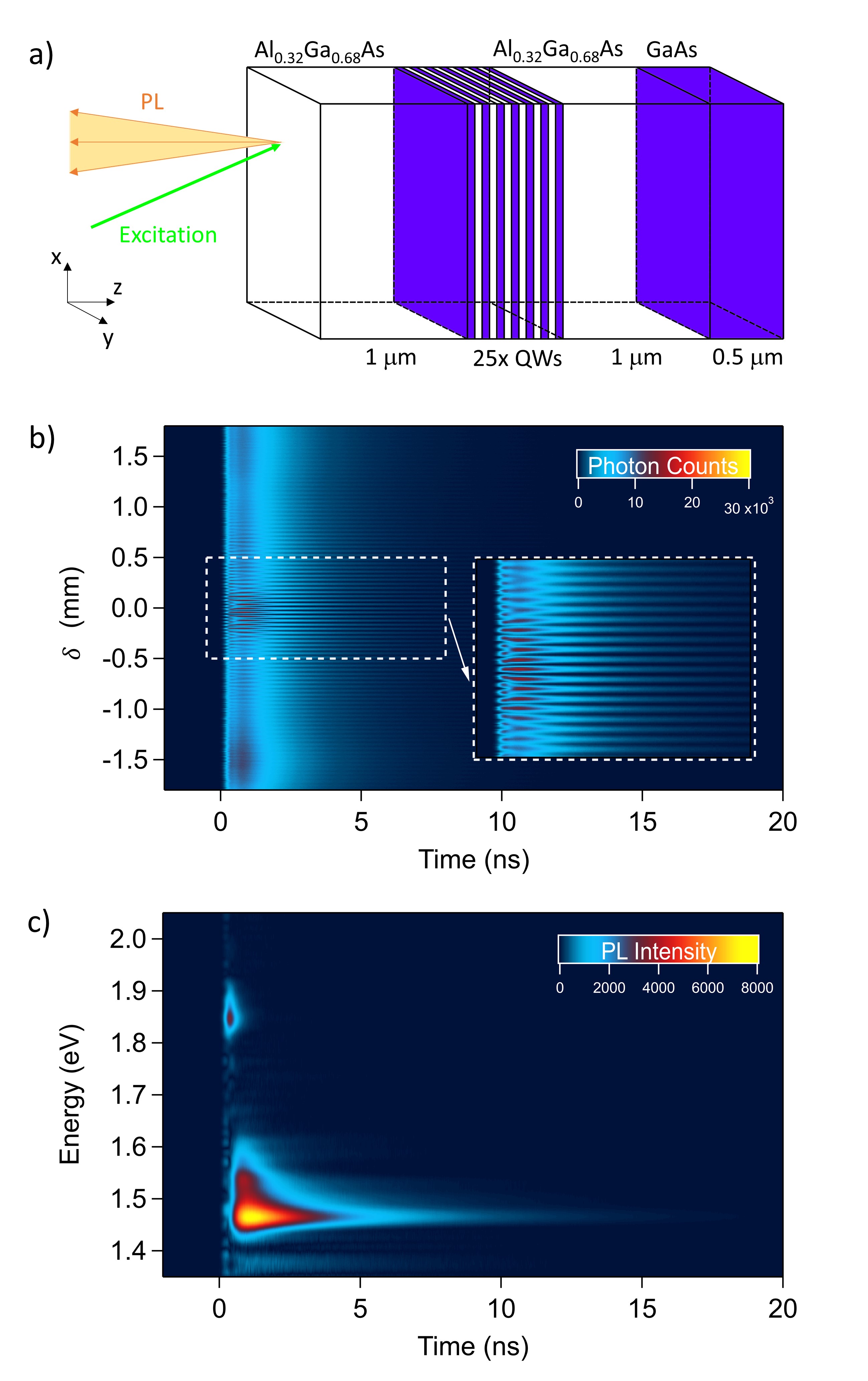}
\caption{\label{fig:Fig2} (a) Schematic of the QW heterostructure and of the excitation-emission geometry. A first 1 $\mu$m-thick layer of Al$_{0.32}$Ga$_{0.68}$As is followed by 25 quantum wells (composed of 120\,\AA\, GaAs quantum well and 150\,\AA\, Al$_{0.32}$Ga$_{0.68}$As barrier layers). The QWs are followed by a second 1 $\mu$m-thick layer of Al$_{0.32}$Ga$_{0.68}$As and by a 0.5 $\mu$m-thick layer of GaAs. (b) Interferograms map, showing the photon counts as a color scale, as a function of time and interferometer wedge position. A zoom is reported as an inset, to evidence the interference pattern. (c) Time-and-spectrally resolved PL map. PL intensity is shown as a color scale versus time and photon energy. This map is obtained by Fourier-transforming each interferogram at fixed time, extracted from the map shown in (b).}
\end{figure}

The sample consists in a sequence of 25- 120 \AA\text{ }GaAs  / 150 \AA\text{ }Al$_{0.32}$Ga$_{0.68}$As MQWs, cladded between two 1 $\mu$m thick Al$_{0.32}$Ga$_{0.68}$As layers (see Fig. 1(a) ). The heterostructure is grown by molecular beam epitaxy (MBE) in standard conditions, at a temperature of 750 °C, on a semi-insulating GaAs(001) wafer after deposition of a 0.5 $\mu$m thick GaAs buffer layer, as schematically shown in Fig. \ref{fig:Fig2}a). The layers were not intentionally doped, the main contaminant is C (p-doping at the given growth conditions) at a concentration of about $10^{14}$ cm$^{-3}$.
\\
Samples are photoexcited by $\approx $ 300 fs laser pulses with a wavelength of 517 nm (2.4 eV), focused on the sample to a 50 $ \mu $m FWHM spot, at a fluence of 33 $ \mu J/cm^2 $ and a repetition rate of 200 kHz. The time-dependent PL signal is acquired via TCSPC (Time-Correlated Single Photon Counting). The spectra in the frequency domain are obtained via a Fourier-Transform (FT) approach\cite{Perri2017,Perri2018}, using a Nireos GEMINI interferometer (see Supplementary Information).
\\
Fig. \ref{fig:Fig2}(b) reports the interferograms map, where the photon counts are displayed as a color scale as a function of time and interferometer wedge position $\delta$. The map is obtained by acquiring histograms (photon counts versus emission time) for each position of the interferometer, scanned from -1.8 mm and 1.8 mm in steps of 5 $\mu$m. These parameters result in a spectral resolution of $\approx$ 16 meV, while the time resolution is $\approx$ 250 ps. \\
Fig. \ref{fig:Fig2}c) shows the time and spectral distribution of the PL signal resulting from the FT of the interferogram extracted at each emission time.
\\
The PL emission is significant only in two spectral regions. The first is centered at a photon energy $h \nu \equiv E \approx$1.85 eV, and lasts for a few hundreds of ps. This emission is assigned to the direct optical gap transitions  of the Al$_{0.32}$Ga$_{0.68}$As layer within the first $1$ $\mu$m thick region. The second, most prominent, covers an energy range between 1.4 eV-1.65 eV. This feature originates from QWs states. Noticeably, its emission maximum is delayed by $>$ $500$ $ps$ with respect to the emission onset. Being the light penetration depth in Al$_{0.32}$Ga$_{0.68}$As $\approx$ 150 nm at 517 nm\cite{RefractiveIndex}, the excitation pulse is almost fully absorbed in this barrier layer near the surface, while only 0.2 \% of laser light can directly excite the QWs. This fact indicates that the QWs are excited indirectly, by the carriers generated in the barrier layer. In particular, most of these excited carriers reach the multiple QWs after thermalization, with a delay due to the hydrodynamic diffusion time of the e-h plasma created by the light pulse in the shallow sub-surface region of the first Al$_{0.32}$Ga$_{0.68}$As barrier layer.
\\
The photoexcited e-h plasma generated near the surface can either recombine non-radiatively through defects at the sample surface, or recombine radiatively in the bulk of Al$_{0.32}$Ga$_{0.68}$As barrier layer, emitting light at $h \nu \approx$1.85 eV, or diffuse to reach the QWs region. The $\approx$500 ps delay between the maxima of the PL intensity from the Al$_{0.32}$Ga$_{0.68}$As barrier layer and from the QWs, which corresponds to the diffusion time of the e-h plasma in the Al$_{0.32}$Ga$_{0.68}$As layer, is much longer than the lifetime of the Al$_{0.32}$Ga$_{0.68}$As luminescence (see Fig. \ref{fig:Fig2}c). Therefore, the decay of the e-h plasma in the barrier layer is not mainly caused by its transfer and injection in the QWs, but by its surface and bulk recombination. The origin of the short lifetime of the PL signal at $\approx$1.85 eV is discussed later in this section.

\begin{figure}
\includegraphics[width=0.9\columnwidth]{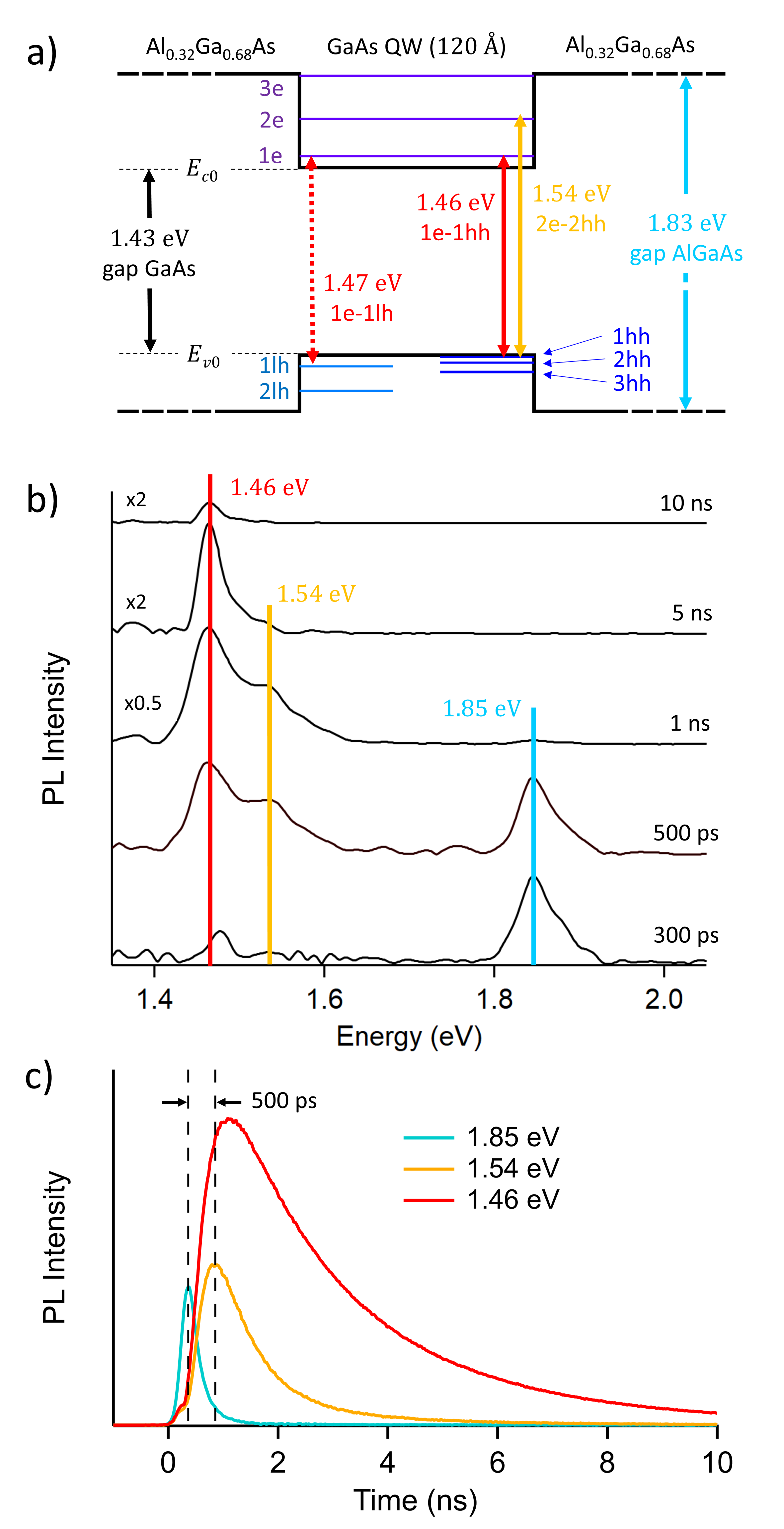}
\caption{\label{fig:Fig3} (a) Schematics of the minima (maxima) of electrons (holes) sub-bands in a Al$_{0.32}$Ga$_{0.68}$As/GaAs/Al$_{0.32}$Ga$_{0.68}$As quantum well. The three lowest-energy-allowed optical transitions are indicated. They are the 1e-1hh (1.46 eV, red), the 1e-1lh (1.47 eV, dashed red), and the 2e-2hh (1.54 eV, yellow). The energy gaps of GaAs (1.43 eV) and Al$_{0.32}$Ga$_{0.68}$As (1.83 eV) are also indicated. (b) PL spectra extracted from the map in Fig. 1(c), at selected delays (indicated on each curve) after excitation. The PL spectra show peaks that can be easily linked to the optical transitions highlighted in panel (a). The energy of each peak as determined experimentally is reported, using the same color code of panel (a). In particular, the 1e-1hh, 2e-2hh, and Al$_{0.32}$Ga$_{0.68}$As gap transitions are detected at energies close to the predicted values, while the emission from the 1e-1lh transition is masked by the broad emission from the 1e-1hh transition. The time evolution of the PL spectrum is discussed in detail in the text; at early delays emission from the Al$_{0.32}$Ga$_{0.68}$As gap is observed, while the population of the QWs states follows. Emission from the lowest-energy QW state persists for the longest time. (c) The PL dynamics at selected energies are reported using the same color code as panels (a) and (b), evidencing the chronology of PL emission.}
\end{figure}

Fig. \ref{fig:Fig3}a) displays the alignment of the energy levels of the heterostructure, at $ k_{\perp} \equiv \sqrt{k_x^2+k_y^2} = 0 $. The energies for electrons ($n$e), light holes ($n$lh) and heavy holes ($n$hh) bands have been calculated by solving the equations for the finite quantum well, using the effective mass and the envelope function approximations. In the following, we indicate with $ E_e^{(n)} $ the energy levels for electrons, referred to the bottom of the bulk GaAs conduction band ($ E_{c0} $) and with $ E_{hh}^{(n)}$ and $ E_{lh}^{(n)} $, the energy levels of heavy and light holes, referred to the top of the bulk GaAs valence band ($ E_{v0} $).
In the present experimental geometry, the lowest-energy allowed transitions are 1e-1hh, 1e-1lh, 2e-2hh, with energies of 1.46 eV, 1.47 eV, 1.54 eV respectively. Higher-energy transitions (not shown) are 2e-2lh (1.60 eV) and 3e-3hh (1.67 eV).
\\
The spectral profiles reported in Fig. \ref{fig:Fig3}b) are extracted at fixed emission time from the PL map of Fig. \ref{fig:Fig2}c). At early delays, the emission at 1.85 eV dominates. Then, the PL emission from the QWs significantly increases. This suggests a retarded growth of the population of the QW excited states. Fig. \ref{fig:Fig3}a) shows the measured emissions related to the 1e-1hh and the 2e-2hh transitions. The 1e-1lh transition (dashed line in fig 2a) is not resolved since its energy separation with the 1e-1hh transition is smaller than the spectral resolution. At longer delays, a persistent PL signal is observed from the lowest-energy transition. This finding suggests the accumulation of electrons and holes in the lowest allowed energy states before their recombination on a nanosecond time scale. Fig. \ref{fig:Fig3}c) shows the PL dynamics in the energy ranges related to the transitions 1e-1hh (red), 2e-2hh (orange) and across the  Al$_{0.32}$Ga$_{0.68}$As gap (light blue), respectively.
\\
The 1e-1hh transition has a first onset at the arrival of the laser pulse and a stronger rise about 0.3 ns after. Since the rise-time of the the 1e-1hh luminescence in similar directly excited QWs is less than 10 ps \cite{DEVEAUD1988435}, we attribute the first weak onset to the direct excitation of the QWs from the few photons not absorbed by tha Al$_{0.32}$Ga$_{0.68}$As layer and the second stronger onset to the arrival in the wells of the plasma excited in the barrier region 0.8-0.9 $\mu$m before the QWs. The carrier capture time into the QW is expected to be of the order of $10^{-11}$ s and do not contribute appreciably to the delay \cite{1999SeScT..14..790M}. 
These data allow to set at $\approx$ 500 ps the delay between the PL emission from the first Al$_{0.32}$Ga$_{0.68}$As barrier layer and from the 1 $\mu$m deeper QWs. Hence the average speed of the electron-hole plasma, while moving from the surface of the Al$_{0.32}$Ga$_{0.68}$As layer to the QWs, is $v_{avg}\approx 2 \cdot$ 10$^5$ cm/s. This value is consistent with the average carrier speed calculated in the diffusive transport regime, using the Al$_{0.32}$Ga$_{0.68}$As diffusion coefficient \cite{IOFFRE}, and considering, slightly after excitation, an e-h plasma density profile along z proportional to the intensity profile of the excitation laser. The time constant of the decay of the 2e-2hh transition is about 1 ns. This value is substantially larger than the 0.3-0.4 ns value measured for states 0.1 eV above the states at minimum energy in similar GaAs QWs at lower plasma density \cite{UCHIKI1985311}. The slower dacay rate of the n=2 states in our data indicates a substantial filling of the states of the n=1 subbands up to at least 0.1 eV above the minimum.
\\
The integrated intensity emitted at 1.85 eV by Al$_{0.32}$Ga$_{0.68}$As is lower than the integrated intensity emitted at 1.54 eV and 1.46 eV by the QWs, despite the fact that only a small fraction of the carriers photo-generated in Al$_{0.32}$Ga$_{0.68}$As enters the QWs. This phenomenon is due to the fast and non-radiative recombination at the sample surface, that strongly quenches the plasma density at the surface of the barrier layer. Hence, only a few PL photons can be emitted from the surface of the barrier layer. Moreover, the radiative lifetime at the initial plasma density near the surface (z<500 nm) of Al$_{0.32}$Ga$_{0.68}$As is of the order 5-50 ps, as calculated using a radiative recombination rate \cite{Modesti1981exp2} of 3.6$\cdot$10$^{-8}$ cm$^3\cdot$s$^{-1}$. This gives rise to a time-resolution-limited PL dynamics at 1.85 eV, with an overall low integrated intensity. Finally, in the barrier region near the QWs, where the initial density is lower than that near the surface, the carrier radiative lifetime increases above 1 ns, however in this case the auto-absorption by the barrier contributes to reduce the intensity of emitted photons. Hence, since radiative recombination is dominant in the QWs, they emit a large amount of PL photons, which can escape the barrier layer without attenuation because the Al$_{0.32}$Ga$_{0.68}$As barrier is transparent to the long-wavelength emission from the QWs.

\section{\label{sec:sec3}Discussion}
\begin{figure*}
\includegraphics[width=\textwidth]{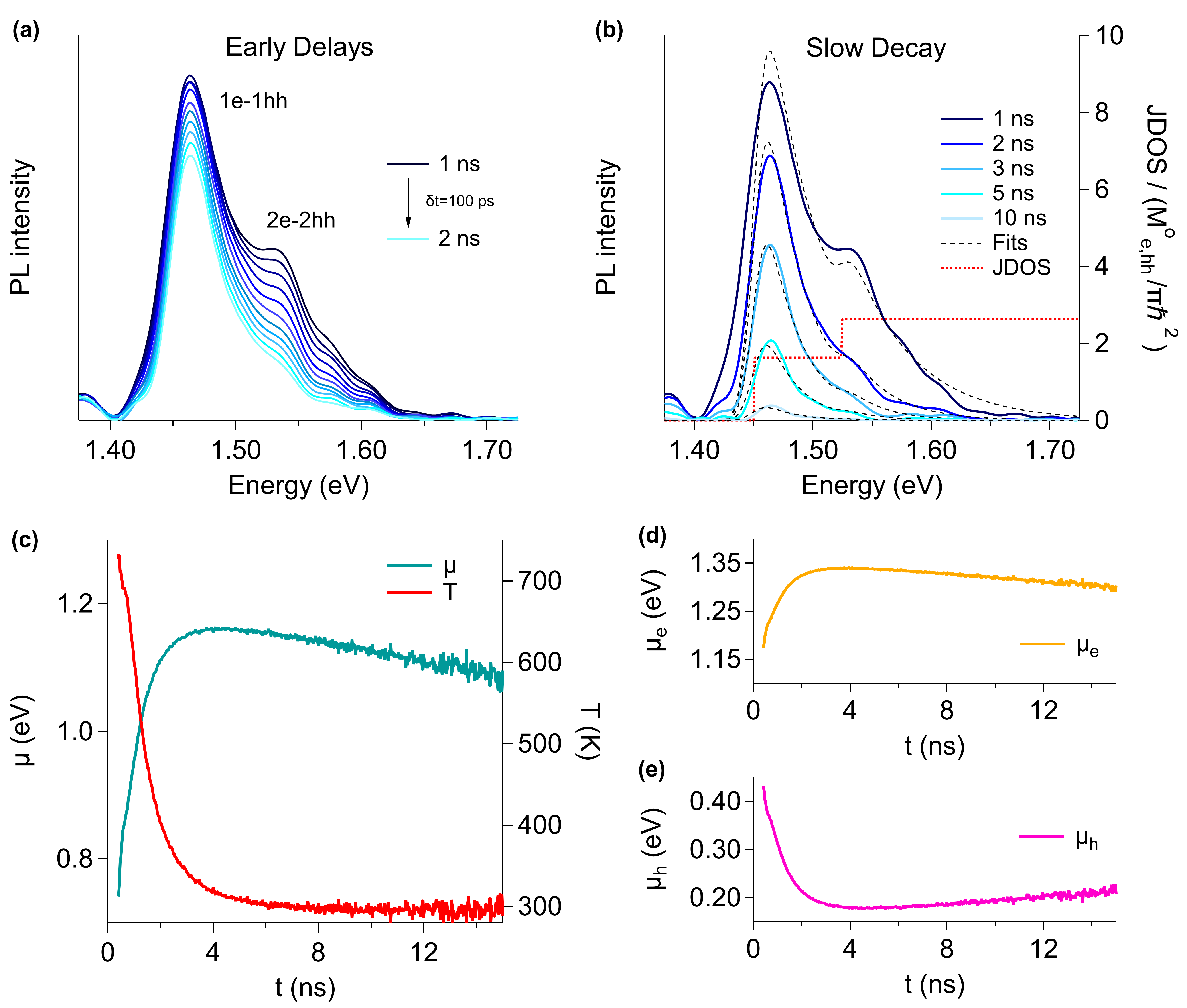}
\caption{\label{fig:Fig4} (a) PL spectra from the QWs 1e-1hh and 2e-2hh optical transitions at early delays, from t=1 ns to t=2 ns in steps of 100 ps. (b) PL spectra at selected delays in the range 1-10 ns. The black dashed lines are the fits to the data, performed as described in the main text. The red dotted line represents the JDOS used in the model, which accounts for the optical transitions 1e-1hh and 1e-1lh (1.45 eV) and 2e-2hh (1.53 eV). (c) Time evolution of the effective chemical potential $\mu$ (light blue curve) and plasma temperature T (red curve) as obtained from the fit. (d)-(e) Time evolution of the chemical potential of electrons ($\mu_e$) and holes ($\mu_h$), computed from $\mu$ as described in the supplementary information.}
\end{figure*}

\begin{figure*}
\includegraphics[width=\textwidth]{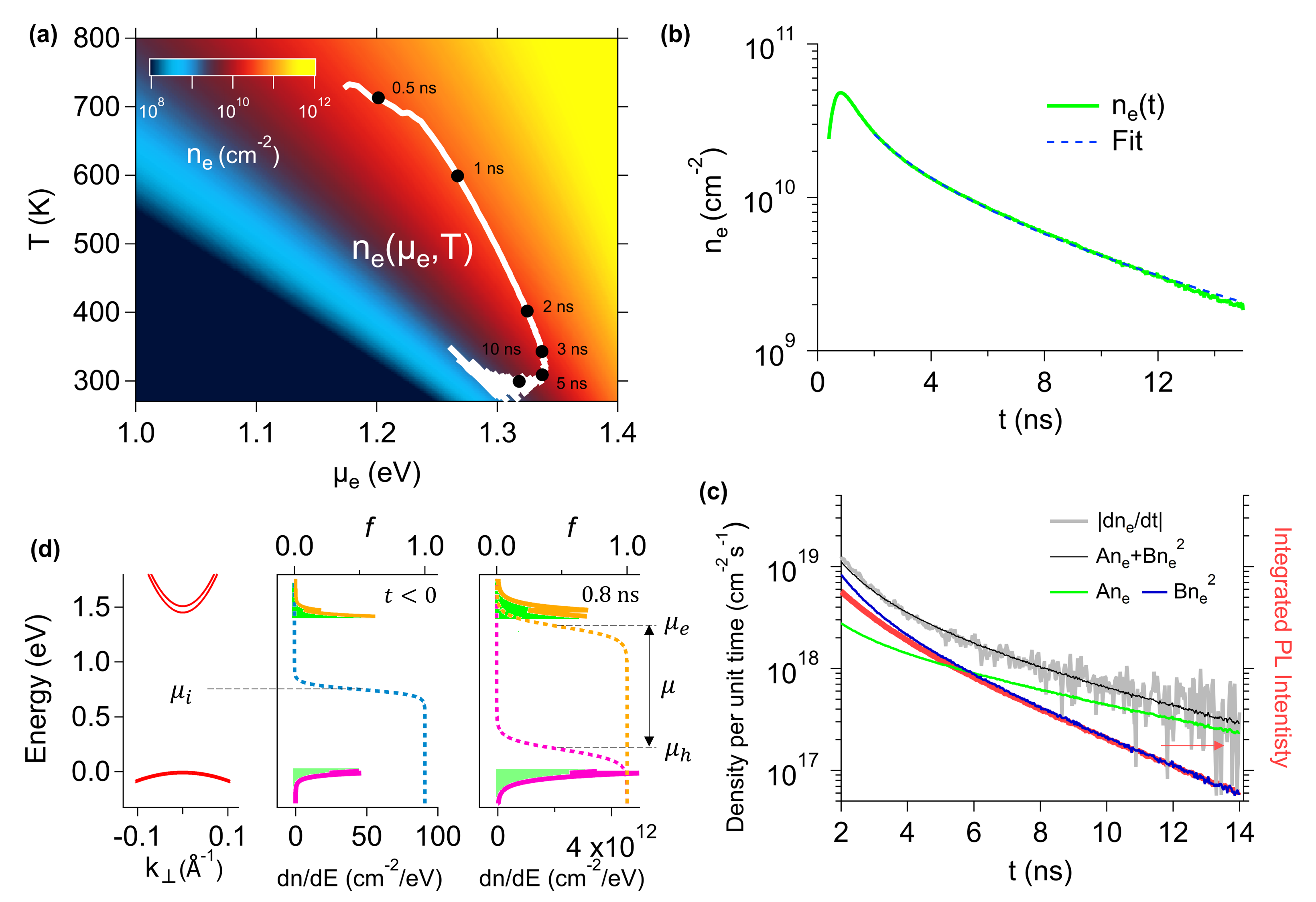}
\caption{\label{fig:Fig5} (a) Bidimensional plot of $n_e(\mu_e,T)$, drawn using Eq. \ref{eq_5} and expressed as a colorscale. The trajectory followed by the system in the ($\mu_e$,T) plane after excitation is marked by the white curve, drawn using $\mu_e$ and T obtained by the fit of PL spectra, and shown in Fig. 3(c). The black dots mark the density reached at selected delays after excitation. (b) Time evolution of the electronic density $n_e$(t)=$n_e(\mu_e$(t)$,T$(t)$)$, computed with Eq. \ref{eq_5} using $\mu_e$ and T from Fig. 4(c). (c) Left axis: terms appearing in the Eq. \ref{eq_6}: $ |dn_e/dt| $ (gray), $ Bn_e^2 $ (blue), $ An_e $ (green), and the sum $ An_e+Bn_e^2 $ (black). This graph shows that at early delays (t$<$3 ns), $ |dn_e/dt| \approx -Bn_e^2 $, hence the recombination is nearly purely bimolecular. Conversely, at late delay (t $\gtrapprox$10 ns), $ |dn_e/dt| \approx -An_e $, hence the recombination is monomolecular. For $ 3 ns < t < 10 ns $, both monomolecular and bimolecular terms are relevant. The crossover $ An_e = Bn_e^2 $ is at $ t \approx 6 $ ns. Right axis: Normalized spectrally-integrated PL intensity (red), which is proportional to $\ n_e^2 $ at late delays, proving that PL arises mainly from the bimolecular recombination of electrons and holes, while the monomolecular decay channel is mostly non-radiative. (d) Sketch of the QWs' band structure around zone-center ($\Gamma$), showing electron and heavy-hole bands (left panel). The middle and right panels show respectively the alignment of the chemical potentials and the carrier density (green areas for electrons, light green areas for holes) before the excitation (t<0) and 0.8 ns after excitation. In the photoexcited system, electrons and holes have their own chemical potential, with $\mu$=$\mu_e$-$\mu_h$.}
\end{figure*}

Assuming that after excitation and diffusion the e-h plasma is thermalized with a constant carrier density distribution in the QWs plane, it is possible to evaluate the occupation of the QWs states using two Fermi-Dirac distributions, for electrons and holes. The depth of the top Al$_{0.32}$Ga$_{0.68}$As layer (1 $\mu$m) is much smaller than the diameter of the laser spot (50 $\mu$m). Therefore the density profile in the x-y plane of the e-h plasma excited in the Al$_{0.32}$Ga$_{0.68}$As layer that reaches the quantum wells should grossly mimik the excitation profile of the Gaussian laser beam in the same plane. If the recombination rate is more than linear in the e-h density (see below) most of the emitted photons come from the dense part of the distribution. In the following we make the approximation of a constant density in the central region of the e-h plasma profile and disregard the contribution of the tails.
\\
Fig. \ref{fig:Fig4}a) shows the PL intensity in the range 1.37-1.75 eV at several time delays (from 1 ns to 2 ns in steps of 100 ps). These curves are extracted from the dataset shown in Fig. \ref{fig:Fig2}c). Under these conditions, the PL spectra are proportional to the joint density of states (JDOS) multiplied by the Fermi Dirac (FD) distribution for electrons ($f_e$) and holes ($f_h$), with a temperature, T, and electrons and holes chemical potentials, $ \mu_e $ and $ \mu_h $, that evolve in time.
\\
Furthermore, in the low excitation limit ($ E_{c0} + E_e^{(1)} - \mu_e >> k_B T $ and $ \mu_h - E_{v0} + E_{hh}^{(1)} >> k_B T $), the product $ f_e \cdot f_h $ can be approximated by a single Boltzmann distribution with an effective chemical potential $ \mu=\mu_e-\mu_h $, which describes the occupation of the joint states (more details are provided in the Supplementary Information).
\\
The resulting fitting function $ I_{PL} (E) $, expressed as a function of the photon energy E, is given by
\begin{equation}\label{eq_1}
    I_{PL}(E) = [ C \cdot e^{-\frac{E-\mu}{k_B T}} \cdot JDOS(E) ] \circledast G
\end{equation}
where G is a Gaussian function that accounts for the finite experimental energy resolution and the emission-line broadening. The JDOS (drawn in Fig. \ref{fig:Fig4}b) )includes only the observed QWs emissions, 1e-1hh/1e-1lh and 2e-2hh, and does not include excitonic effects. The constant C aims to match the intensity of the PL signal, and the way it was evaluated is given in the Supplementary Information.
\\
The fits to the PL spectra are reported in Fig. 3b for a few representative time delays. In this model, the only free fitting parameters used to match the main features of the QWs PL at all emission times are T and $\mu$. 
In the fits, the Gaussian FWHM is set to 20 meV. Since the JDOS is fixed, the effect of the density-dependent band-gap renormalization and all the possible many-body interactions of the excited plasma are neglected. The band gap renormalization of the e-h plasma in GaAs quantum wells\cite{PhysRevLett.58.419} is below 20 meV for densities below $10^{11}$ cm$^{-2}$, i.e. is comparable to or lower than our energy resolution for the e-h densities we obtain (see below). For this reason we neglect the renormalization, toghether with the effects of other many body interactions. The constant C is also kept fixed. We note that, at early time-delays, where the plasma density is maximal, there exists a discrepancy between the spectra and the fit in the gap-energy-region of GaAs (1.43 eV). Likely this is the joint effect of the gap renormalization, the many body broadening and the density gradient which show up at the highest density \cite{Trankle1987-gs}. The same explanation holds for the smoothing of the shoulder at 1.54 eV.
\\
The JDOS is reported in Fig. \ref{fig:Fig4}b) as a dashed line. It is plotted in units of $M^o_{e,hh}/\pi\hbar^2$, where $ M^o_{e,hh} $ is the optical mass of electrons - heavy holes optical transitions.
\\
In the JDOS, the contribution of 1e-1lh transition is included by considering the first step height as the sum of heavy and light holes optical masses.
\\
Fig. \ref{fig:Fig4}c) shows T and $\mu$ versus time, as obtained from the fit of the PL spectra. The system returns to room temperature ($ \approx 300 K $) with an exponential time constant of $\approx$ 1 ns. Conversely, after a fast increase, the chemical potential decays monotonically with time. The initial high temperature of carriers in the QWs is caused by the onset of the photoexcitation and by the delayed injection of the photo-generated hot plasma from the surface layer. The dynamics on the nanosecond timescale reflect the cooling of the crystal lattice \cite{DiCicco2020}.
The fact that the PL intensity of the 2e-2hh transition decays faster than the 1e-1hh transition is due to the decrease of the temperature that reduces the occupation of the higher energy states.
\\
The knowledge of temperature and chemical potential is the prerequisite for the estimation of the carrier density. However, it is first necessary to compute $\mu_e$ and $\mu_h$ starting from $\mu$ as obtained by the fits (see the Supplementary Information). Fig. \ref{fig:Fig4}d) displays how $\mu_e$ and $\mu_h$ evolve in time.
\\
In the limit $ E_{c0} + E_e^{(1)} - \mu_e >> k_B T $, and considering only the first and the second energy levels for electrons (see Fig. \ref{fig:Fig3}a) ), the electronic density is (see Supplementary Information):
\begin{equation}\label{eq_5}
    n_e(T,\mu_e) \approx \frac{m_e^*}{ \pi \hbar^2} k_B T e^{-\frac{E_{c0}+E_e^{(1)}-\mu_e}{k_B T} } (1+e^{-\frac{E_e^{(2)}-E_e^{(1)}}{k_B T} })
    \end{equation}
The 2D map reported in Fig. \ref{fig:Fig5}a) is obtained by using this equation. The electron density in the QWs (in units of $cm^{-2}$) is plotted versus $ \mu_e $ and T. The white curve $( \mu_e,T )$ marks the time evolution of the electron's density. At t$\approx$0.5 ns, the electronic chemical potential and the temperature are $\approx$1.20 eV and $\approx$710 K respectively, as determined from the fitting procedure (see Fig. \ref{fig:Fig4}c)). Then, the temperature decreases from $\approx$710 K to a constant value of $\approx$300 K within $\approx$4 ns. The slow e-h plasma cooling is determined by the competition between carrier-phonon scattering (fast cooling) in the QWs and injection of hot free electrons from Al$_{0.32}$Ga$_{0.68}$As (heat source).
The chemical potential first increases, in the time interval t$\approx$0.5-3 ns, then it decreases monotonically.
\\
By using the time-varying T and $\mu_e$ values and Eq. \ref{eq_5}, $n_e$(t) is obtained. The results are reported in Fig. \ref{fig:Fig5}b). $n_e$(t) increases in the interval 0.4-0.8 ns because of carriers injection from the Al$_{0.32}$Ga$_{0.68}$ layer, that overcomes the effect of e-h recombination. The maximum density, found at t=0.8 ns, is equal to n$_{e,max}\approx$5$\cdot$10$^{10}$ cm$^{-2}$, corresponding to a volume density of $\approx$4$\cdot$10$^{16}$ cm$^{-3}$. This value is calculated assuming a 1 $\mu$m-thick Al$_{0.32}$Ga$_{0.68}$As barrier layer, with an absorption coefficient equal to 63840 cm$^{-1}$ at 515 nm \cite{RefractiveIndex}. For t>0.8 ns, $n_e$(t) decreases monotonically.
\\
The time evolution of $n_e$ provides information about the mechanism responsible for electron-hole recombination. The rate equation for the electron density is given by \cite{Xing2017}:
\begin{equation}\label{eq_6}
	\frac{dn_e(t)}{dt} = G(t) - An_e(t) - Bn_e^2(t)
\end{equation}
where the term $-An_e$ accounts for mono-molecular and non-radiative e-h recombination, $-Bn_e^2$ is the bi-molecular radiative recombination term for the e-h plasma, and G(t) is the e-h generation term (optical excitation or free carrier injection). In our experimental conditions, at room temperature, the excitonic contribution, which is monomolecular and radiative, is negligible. Once the plasma injection from Al$_{0.32}$Ga$_{0.68}$As is over (t $\gtrapprox$ 2 ns), we can set $G$(t)=0. Using this assumption, the solution of Eq. \ref{eq_6} gives:
\begin{equation}
	n_e(t)=\frac{A/B}{e^{A(t-t_0)+k}-1}
\end{equation}
with $ t_0=2\cdot10^{-9} $ s, and where k is a constant set by the initial value of $n_e$ (i.e. $n_e$(t=2 ns) in the present case). This function fits the electronic density in the interval 2-15 ns. The fit curve is plotted as a dashed blue line in figure \ref{fig:Fig5}b). The fitting parameters are: A=1.05$\cdot 10^{8} \pm 8 \cdot 10^{6} $ s$^{-1}$, B=1.20 $\cdot 10^{-2} \pm 8 \cdot 10^{-4} $ cm$^{2}$ s$^{-1}$, and $ k=0.29 \pm 0.01 $. The A parameter corresponds to a non-radiative lifetime of $ 10^{-8} $ s. This is very short compared to the expected non-radiative lifetime due only to the residual doping (p type $10^{14}$ $cm^{-3}$ our sample). As a reference \cite{ILAHI2023414612}, a non radiative lifetime of $ 10^{-7} $ s is found for a GaAs QW with a residual doping level of $ 5 * 10^{17} $ $cm^{-3}$ . This fact suggests a dominant contribution of traps at the AlGaAs-GaAs interfaces \cite{1073154}. The value of the B parameter is consistent with previous investigations on GaAs/AlGaAs QW \cite{Zhang2022}.
\\
Fig. \ref{fig:Fig5}c) reports the terms of Eq. \ref{eq_6}, where A and B are the coefficients obtained from the fit of $ n_e(t) $. These data show that at early delays ($\approx$2-6 ns) the density decay is mainly due to the bimolecular term $ -Bn_e^2$, suggesting that the recombination of free electrons and holes is the dominant process. As the delay increases, the monomolecular term $-An_e $ becomes more relevant. The crossover between bimolecular and monomolecular regime is found to be at $t \approx $ 6 ns. Furthermore, we observe that the spectrally-integrated PL signal, $ I_{PL}(t)$, is closely proportional to $ n_e^2$. This fact suggests that: i) PL is mainly due to the bimolecular recombination of free electrons and holes, while the monomolecular excitons contribution is negligible; ii) the monomolecular decay channel is mostly non-radiative. We associate this recombination channel to the capture of carriers by traps and defects (Shockley Read Hall). At high carrier density, the traps are saturated, and their contribution to the recombination rate is negligible.
\\
Fig. \ref{fig:Fig5}d) shows a simplified scheme for the occupation of the electronic states in an intrinsic semiconductor, before and after the generation of an electron-hole plasma via direct or indirect excitation. The energy levels relevant to the present case are sketched in the left panel. At equilibrium (t<0, middle panel), the chemical potential assumes its intrinsic value $\mu_i$ and only a few electrons occupy the conduction band. Their density $ n_e $ is represented by the green area. At 0.8 ns after excitation (right panel) a large number of electrons are promoted in the conduction band. As a consequence, the chemical potential splits for electrons and holes, while their densities increase significantly.
\\
Finally, it is worth noticing that the structure of the present sample allows decoupling the photoexcitation mechanism from the filling of the QW states. In particular, the excited carriers reach the GaAs QWs after thermalization in the Al$_{0.32}$Ga$_{0.68}$As barrier layer. Therefore, they enter the QWs with an excess energy of $\approx$0.4 eV, corresponding to the energy difference of the Al$_{0.32}$Ga$_{0.68}$As and GaAs bandgaps. In the case of a direct excitation, the excess energy would have been larger, of the order $\approx$0.94 eV, corresponding to the energy difference between the excitation photon energy $h \nu$=2.4 eV and the GaAs bandgap. Hence, since carrier thermalization occurs $\approx$1 $\mu$m away from the QWs in the case of indirect excitation, the temperature rise of the QWs is low. As discussed before, despite surface and bulk recombination of the plasma in the barrier layer, in the present QWs heterostructure the amount of carriers transferred from the barrier layer to the QWs is larger than that of the carriers directly excited in the QW. This fact suggests that an indirect excitation from a thick barrier layer can provide a denser and cooler plasma with respect to the effect of a direct photoexcitation of the QWs, together with a slower injection of the already generated plasma. Finally, a way to enhance the fraction of indirectly generated plasma available to fill the QWs would be to engineer the layers composing the heterostructure. As an example, a 10s of nm thin additional layer with energy gap larger than that of the barrier layer (achieved by a larger Al concentration than in the barrier layer), placed at the sample surface, would prevent surface recombination, hence increasing the effective plasma density available in the QWs.

\section{\label{sec:sec4}Conclusions\protect}
In this work, we measured the time and spectral characteristics of the photoluminescence from a complex QW heterostructure. By accessing the temporal dynamics in the sub-ns domain, the chronology of these emissions, with the PL from the Al$_{0.32}$Ga$_{0.68}$As band-gap being emitted instantaneously is unveiled. In particular, the PL from GaAs QW's, located at 1 $\mu$m depth from the heterostructure surface, is delayed by $\approx$500 ps, signaling that the population of the QW subbands is obtained mainly upon diffusion of the electron-hole plasma from the Al$_{0.32}$Ga$_{0.68}$As layer. This delay is connected to the diffusion speed of the electron-hole plasma in the Al$_{0.32}$Ga$_{0.68}$As layer.
We successfully modeled the time-evolution of the PL spectrum and proved that the filling of the QWs states can be completely described by two Fermi-Dirac distributions with temperature T and chemical potentials $\mu_e$ and $\mu_h$ evolving in time. We showed that the time-dependent PL emission from the QWs is linked to the time evolution of the electron density and entirely governed by the time evolution of temperature and chemical potentials. The electrons/holes carrier densities decay mostly because of bimolecular radiative recombination, although at large delays (low density) a mono-molecular non-radiative process dominates.
\\
In conclusion, the combined experimental access to both spectral and temporal domains allows us to discern the characteristics of light emission from complex artificial heterostructures. This approach can thus provide useful information for characterizing engineered nanoscale structures and achieving spectral features and dynamics tailored to specific luminescence devices applications, in which both the color of the emitted light and the instant of emission might be crucial for reaching new functionalities.

\section*{Data Availability Statement}
The data that support the findings of this study are available from the corresponding author upon reasonable request.

\section*{Disclosures}
The authors declare no conflicts of interest.

\section*{Acknowledgements}
The authors thank Prof. Silvio Modesti for providing the sample and for valuable discussion about data analysis and interpretation of the results.

\section*{References}
\bibliography{aipsamp}

\begin{thebibliography}{38}%
\makeatletter
\providecommand \@ifxundefined [1]{%
 \@ifx{#1\undefined}
}%
\providecommand \@ifnum [1]{%
 \ifnum #1\expandafter \@firstoftwo
 \else \expandafter \@secondoftwo
 \fi
}%
\providecommand \@ifx [1]{%
 \ifx #1\expandafter \@firstoftwo
 \else \expandafter \@secondoftwo
 \fi
}%
\providecommand \natexlab [1]{#1}%
\providecommand \enquote  [1]{``#1''}%
\providecommand \bibnamefont  [1]{#1}%
\providecommand \bibfnamefont [1]{#1}%
\providecommand \citenamefont [1]{#1}%
\providecommand \href@noop [0]{\@secondoftwo}%
\providecommand \href [0]{\begingroup \@sanitize@url \@href}%
\providecommand \@href[1]{\@@startlink{#1}\@@href}%
\providecommand \@@href[1]{\endgroup#1\@@endlink}%
\providecommand \@sanitize@url [0]{\catcode `\\12\catcode `\$12\catcode `\&12\catcode `\#12\catcode `\^12\catcode `\_12\catcode `\%12\relax}%
\providecommand \@@startlink[1]{}%
\providecommand \@@endlink[0]{}%
\providecommand \url  [0]{\begingroup\@sanitize@url \@url }%
\providecommand \@url [1]{\endgroup\@href {#1}{\urlprefix }}%
\providecommand \urlprefix  [0]{URL }%
\providecommand \Eprint [0]{\href }%
\providecommand \doibase [0]{http://dx.doi.org/}%
\providecommand \selectlanguage [0]{\@gobble}%
\providecommand \bibinfo  [0]{\@secondoftwo}%
\providecommand \bibfield  [0]{\@secondoftwo}%
\providecommand \translation [1]{[#1]}%
\providecommand \BibitemOpen [0]{}%
\providecommand \bibitemStop [0]{}%
\providecommand \bibitemNoStop [0]{.\EOS\space}%
\providecommand \EOS [0]{\spacefactor3000\relax}%
\providecommand \BibitemShut  [1]{\csname bibitem#1\endcsname}%
\let\auto@bib@innerbib\@empty
\bibitem [{\citenamefont {McHale}(2017)}]{McHale}%
  \BibitemOpen
  \bibfield  {author} {\bibinfo {author} {\bibfnamefont {J.~L.}\ \bibnamefont {McHale}},\ }\bibfield  {title} {\enquote {\bibinfo {title} {Molecular spectroscopy},}\ }in\ \href {\doibase https://doi.org/10.1201/9781315115214} {\emph {\bibinfo {booktitle} {Molecular Spectroscopy}}},\ \bibinfo {editor} {edited by\ \bibinfo {editor} {\bibfnamefont {J.~L.}\ \bibnamefont {McHale}}}\ (\bibinfo  {publisher} {Taylor \& Francis},\ \bibinfo {year} {2017})\ p.\ \bibinfo {pages} {475}\BibitemShut {NoStop}%
\bibitem [{\citenamefont {Leah~Bergman}(2012)}]{Bergman}%
  \BibitemOpen
  \bibfield  {author} {\bibinfo {author} {\bibfnamefont {J.~L.~M.}\ \bibnamefont {Leah~Bergman}},\ }\bibfield  {title} {\enquote {\bibinfo {title} {Handbook of luminescent semiconductor materials},}\ }in\ \href {\doibase https://doi.org/10.1201/b11201} {\emph {\bibinfo {booktitle} {Handbook of Luminescent Semiconductor Materials}}},\ \bibinfo {editor} {edited by\ \bibinfo {editor} {\bibfnamefont {J.~L.~M.}\ \bibnamefont {Leah~Bergman}}}\ (\bibinfo  {publisher} {Taylor \& Francis},\ \bibinfo {year} {2012})\ p.\ \bibinfo {pages} {468}\BibitemShut {NoStop}%
\bibitem [{\citenamefont {Lakowicz}(2007)}]{Lakowicz}%
  \BibitemOpen
  \bibfield  {author} {\bibinfo {author} {\bibfnamefont {J.~R.}\ \bibnamefont {Lakowicz}},\ }\bibfield  {title} {\enquote {\bibinfo {title} {Principles of fluorescence spectroscopy},}\ }in\ \href {\doibase https://doi.org/10.1007/978-0-387-46312-4} {\emph {\bibinfo {booktitle} {Principles of Fluorescence Spectroscopy}}},\ \bibinfo {editor} {edited by\ \bibinfo {editor} {\bibfnamefont {J.~R.}\ \bibnamefont {Lakowicz}}}\ (\bibinfo  {publisher} {Springer, Boston, MA},\ \bibinfo {year} {2007})\ p.\ \bibinfo {pages} {954}\BibitemShut {NoStop}%
\bibitem [{\citenamefont {Cubeddu}\ \emph {et~al.}(2002)\citenamefont {Cubeddu}, \citenamefont {Comelli}, \citenamefont {D{\textquotesingle}Andrea}, \citenamefont {Taroni},\ and\ \citenamefont {Valentini}}]{Cubeddu2002}%
  \BibitemOpen
  \bibfield  {author} {\bibinfo {author} {\bibfnamefont {R.}~\bibnamefont {Cubeddu}}, \bibinfo {author} {\bibfnamefont {D.}~\bibnamefont {Comelli}}, \bibinfo {author} {\bibfnamefont {C.}~\bibnamefont {D{\textquotesingle}Andrea}}, \bibinfo {author} {\bibfnamefont {P.}~\bibnamefont {Taroni}}, \ and\ \bibinfo {author} {\bibfnamefont {G.}~\bibnamefont {Valentini}},\ }\bibfield  {title} {\enquote {\bibinfo {title} {Time-resolved fluorescence imaging in biology and medicine},}\ }\href {\doibase 10.1088/0022-3727/35/9/201} {\bibfield  {journal} {\bibinfo  {journal} {Journal of Physics D: Applied Physics}\ }\textbf {\bibinfo {volume} {35}},\ \bibinfo {pages} {R61--R76} (\bibinfo {year} {2002})}\BibitemShut {NoStop}%
\bibitem [{\citenamefont {Christensen}\ \emph {et~al.}(2006)\citenamefont {Christensen}, \citenamefont {N{\o}rgaard}, \citenamefont {Bro},\ and\ \citenamefont {Engelsen}}]{Christensen2006}%
  \BibitemOpen
  \bibfield  {author} {\bibinfo {author} {\bibfnamefont {J.}~\bibnamefont {Christensen}}, \bibinfo {author} {\bibfnamefont {L.}~\bibnamefont {N{\o}rgaard}}, \bibinfo {author} {\bibfnamefont {R.}~\bibnamefont {Bro}}, \ and\ \bibinfo {author} {\bibfnamefont {S.~B.}\ \bibnamefont {Engelsen}},\ }\bibfield  {title} {\enquote {\bibinfo {title} {Multivariate autofluorescence of intact food systems},}\ }\href {\doibase 10.1021/cr050019q} {\bibfield  {journal} {\bibinfo  {journal} {Chemical Reviews}\ }\textbf {\bibinfo {volume} {106}},\ \bibinfo {pages} {1979--1994} (\bibinfo {year} {2006})}\BibitemShut {NoStop}%
\bibitem [{\citenamefont {Piesch}, \citenamefont {Burgkhardt},\ and\ \citenamefont {Vilgis~(INVITED)}(1990)}]{Piesch1990}%
  \BibitemOpen
  \bibfield  {author} {\bibinfo {author} {\bibfnamefont {E.}~\bibnamefont {Piesch}}, \bibinfo {author} {\bibfnamefont {B.}~\bibnamefont {Burgkhardt}}, \ and\ \bibinfo {author} {\bibfnamefont {M.}~\bibnamefont {Vilgis~(INVITED)}},\ }\bibfield  {title} {\enquote {\bibinfo {title} {{Photoluminescence Dosimetry: Progress and Present State of Art}},}\ }\href {\doibase 10.1093/oxfordjournals.rpd.a080796} {\bibfield  {journal} {\bibinfo  {journal} {Radiation Protection Dosimetry}\ }\textbf {\bibinfo {volume} {33}},\ \bibinfo {pages} {215--226} (\bibinfo {year} {1990})}\BibitemShut {NoStop}%
\bibitem [{\citenamefont {Wagni{\`e}res}, \citenamefont {Star},\ and\ \citenamefont {Wilson}(1998)}]{Wagnieres1998}%
  \BibitemOpen
  \bibfield  {author} {\bibinfo {author} {\bibfnamefont {G.~A.}\ \bibnamefont {Wagni{\`e}res}}, \bibinfo {author} {\bibfnamefont {W.~M.}\ \bibnamefont {Star}}, \ and\ \bibinfo {author} {\bibfnamefont {B.~C.}\ \bibnamefont {Wilson}},\ }\bibfield  {title} {\enquote {\bibinfo {title} {In vivo fluorescence spectroscopy and imaging for oncological applications},}\ }\href@noop {} {\bibfield  {journal} {\bibinfo  {journal} {Photochem Photobiol}\ }\textbf {\bibinfo {volume} {68}},\ \bibinfo {pages} {603--632} (\bibinfo {year} {1998})}\BibitemShut {NoStop}%
\bibitem [{\citenamefont {Klingshirn}(2007)}]{Klingshirn}%
  \BibitemOpen
  \bibfield  {author} {\bibinfo {author} {\bibfnamefont {C.~F.}\ \bibnamefont {Klingshirn}},\ }\bibfield  {title} {\enquote {\bibinfo {title} {Semiconductor optics},}\ }in\ \href {\doibase https://doi.org/10.1007/b138175} {\emph {\bibinfo {booktitle} {Semiconductor Optics}}},\ \bibinfo {editor} {edited by\ \bibinfo {editor} {\bibfnamefont {C.~F.}\ \bibnamefont {Klingshirn}}}\ (\bibinfo  {publisher} {Springer, Berlin, Heidelberg},\ \bibinfo {year} {2007})\ p.\ \bibinfo {pages} {797}\BibitemShut {NoStop}%
\bibitem [{\citenamefont {Wieghold}\ \emph {et~al.}(2020{\natexlab{a}})\citenamefont {Wieghold}, \citenamefont {Bieber}, \citenamefont {Mardani}, \citenamefont {Siegrist},\ and\ \citenamefont {Nienhaus}}]{Wieghold2020}%
  \BibitemOpen
  \bibfield  {author} {\bibinfo {author} {\bibfnamefont {S.}~\bibnamefont {Wieghold}}, \bibinfo {author} {\bibfnamefont {A.~S.}\ \bibnamefont {Bieber}}, \bibinfo {author} {\bibfnamefont {M.}~\bibnamefont {Mardani}}, \bibinfo {author} {\bibfnamefont {T.}~\bibnamefont {Siegrist}}, \ and\ \bibinfo {author} {\bibfnamefont {L.}~\bibnamefont {Nienhaus}},\ }\bibfield  {title} {\enquote {\bibinfo {title} {Understanding the effect of light and temperature on the optical properties and stability of mixed-ion halide perovskites},}\ }\href {\doibase 10.1039/D0TC02103B} {\bibfield  {journal} {\bibinfo  {journal} {J. Mater. Chem. C}\ }\textbf {\bibinfo {volume} {8}},\ \bibinfo {pages} {9714--9723} (\bibinfo {year} {2020}{\natexlab{a}})}\BibitemShut {NoStop}%
\bibitem [{\citenamefont {Wieghold}\ \emph {et~al.}(2020{\natexlab{b}})\citenamefont {Wieghold}, \citenamefont {Bieber}, \citenamefont {VanOrman}, \citenamefont {Rodriguez},\ and\ \citenamefont {Nienhaus}}]{Wieghold2020b}%
  \BibitemOpen
  \bibfield  {author} {\bibinfo {author} {\bibfnamefont {S.}~\bibnamefont {Wieghold}}, \bibinfo {author} {\bibfnamefont {A.~S.}\ \bibnamefont {Bieber}}, \bibinfo {author} {\bibfnamefont {Z.~A.}\ \bibnamefont {VanOrman}}, \bibinfo {author} {\bibfnamefont {A.}~\bibnamefont {Rodriguez}}, \ and\ \bibinfo {author} {\bibfnamefont {L.}~\bibnamefont {Nienhaus}},\ }\bibfield  {title} {\enquote {\bibinfo {title} {Is disorder beneficial in perovskite-sensitized solid-state upconversion? the role of dbp doping in rubrene},}\ }\href {\doibase 10.1021/acs.jpcc.0c05290} {\bibfield  {journal} {\bibinfo  {journal} {The Journal of Physical Chemistry C}\ }\textbf {\bibinfo {volume} {124}},\ \bibinfo {pages} {18132--18140} (\bibinfo {year} {2020}{\natexlab{b}})}\BibitemShut {NoStop}%
\bibitem [{\citenamefont {Franceschini}\ \emph {et~al.}(2020)\citenamefont {Franceschini}, \citenamefont {Carletti}, \citenamefont {Pushkarev}, \citenamefont {Preda}, \citenamefont {Perri}, \citenamefont {Tognazzi}, \citenamefont {Ronchi}, \citenamefont {Ferrini}, \citenamefont {Pagliara}, \citenamefont {Banfi}, \citenamefont {Polli}, \citenamefont {Cerullo}, \citenamefont {De~Angelis}, \citenamefont {Makarov},\ and\ \citenamefont {Giannetti}}]{Franceschini2020}%
  \BibitemOpen
  \bibfield  {author} {\bibinfo {author} {\bibfnamefont {P.}~\bibnamefont {Franceschini}}, \bibinfo {author} {\bibfnamefont {L.}~\bibnamefont {Carletti}}, \bibinfo {author} {\bibfnamefont {A.~P.}\ \bibnamefont {Pushkarev}}, \bibinfo {author} {\bibfnamefont {F.}~\bibnamefont {Preda}}, \bibinfo {author} {\bibfnamefont {A.}~\bibnamefont {Perri}}, \bibinfo {author} {\bibfnamefont {A.}~\bibnamefont {Tognazzi}}, \bibinfo {author} {\bibfnamefont {A.}~\bibnamefont {Ronchi}}, \bibinfo {author} {\bibfnamefont {G.}~\bibnamefont {Ferrini}}, \bibinfo {author} {\bibfnamefont {S.}~\bibnamefont {Pagliara}}, \bibinfo {author} {\bibfnamefont {F.}~\bibnamefont {Banfi}}, \bibinfo {author} {\bibfnamefont {D.}~\bibnamefont {Polli}}, \bibinfo {author} {\bibfnamefont {G.}~\bibnamefont {Cerullo}}, \bibinfo {author} {\bibfnamefont {C.}~\bibnamefont {De~Angelis}}, \bibinfo {author} {\bibfnamefont {S.~V.}\ \bibnamefont {Makarov}}, \ and\ \bibinfo {author} {\bibfnamefont {C.}~\bibnamefont {Giannetti}},\ }\bibfield  {title} {\enquote
  {\bibinfo {title} {Tuning the ultrafast response of fano resonances in halide perovskite nanoparticles},}\ }\href {\doibase 10.1021/acsnano.0c05710} {\bibfield  {journal} {\bibinfo  {journal} {ACS Nano}\ }\textbf {\bibinfo {volume} {14}},\ \bibinfo {pages} {13602--13610} (\bibinfo {year} {2020})}\BibitemShut {NoStop}%
\bibitem [{\citenamefont {Manzeli}\ \emph {et~al.}(2017)\citenamefont {Manzeli}, \citenamefont {Ovchinnikov}, \citenamefont {Pasquier}, \citenamefont {Yazyev},\ and\ \citenamefont {Kis}}]{Manzeli2017}%
  \BibitemOpen
  \bibfield  {author} {\bibinfo {author} {\bibfnamefont {S.}~\bibnamefont {Manzeli}}, \bibinfo {author} {\bibfnamefont {D.}~\bibnamefont {Ovchinnikov}}, \bibinfo {author} {\bibfnamefont {D.}~\bibnamefont {Pasquier}}, \bibinfo {author} {\bibfnamefont {O.~V.}\ \bibnamefont {Yazyev}}, \ and\ \bibinfo {author} {\bibfnamefont {A.}~\bibnamefont {Kis}},\ }\bibfield  {title} {\enquote {\bibinfo {title} {2d transition metal dichalcogenides},}\ }\href {\doibase 10.1038/natrevmats.2017.33} {\bibfield  {journal} {\bibinfo  {journal} {Nature Reviews Materials}\ }\textbf {\bibinfo {volume} {2}},\ \bibinfo {pages} {17033} (\bibinfo {year} {2017})}\BibitemShut {NoStop}%
\bibitem [{\citenamefont {Tran}, \citenamefont {Kim},\ and\ \citenamefont {Lee}(2016)}]{Tran2016}%
  \BibitemOpen
  \bibfield  {author} {\bibinfo {author} {\bibfnamefont {M.~D.}\ \bibnamefont {Tran}}, \bibinfo {author} {\bibfnamefont {J.-H.}\ \bibnamefont {Kim}}, \ and\ \bibinfo {author} {\bibfnamefont {Y.~H.}\ \bibnamefont {Lee}},\ }\bibfield  {title} {\enquote {\bibinfo {title} {Tailoring photoluminescence of monolayer transition metal dichalcogenides},}\ }\href {\doibase https://doi.org/10.1016/j.cap.2016.03.023} {\bibfield  {journal} {\bibinfo  {journal} {Current Applied Physics}\ }\textbf {\bibinfo {volume} {16}},\ \bibinfo {pages} {1159--1174} (\bibinfo {year} {2016})},\ \bibinfo {note} {special Section on Nanostructure Physics and Materials Science at Center for Integrated Nanostructure Physics, Institute for Basic Science at Sungkyunkwan University}\BibitemShut {NoStop}%
\bibitem [{\citenamefont {Brem}\ \emph {et~al.}(2020)\citenamefont {Brem}, \citenamefont {Ekman}, \citenamefont {Christiansen}, \citenamefont {Katsch}, \citenamefont {Selig}, \citenamefont {Robert}, \citenamefont {Marie}, \citenamefont {Urbaszek}, \citenamefont {Knorr},\ and\ \citenamefont {Malic}}]{Brem2020}%
  \BibitemOpen
  \bibfield  {author} {\bibinfo {author} {\bibfnamefont {S.}~\bibnamefont {Brem}}, \bibinfo {author} {\bibfnamefont {A.}~\bibnamefont {Ekman}}, \bibinfo {author} {\bibfnamefont {D.}~\bibnamefont {Christiansen}}, \bibinfo {author} {\bibfnamefont {F.}~\bibnamefont {Katsch}}, \bibinfo {author} {\bibfnamefont {M.}~\bibnamefont {Selig}}, \bibinfo {author} {\bibfnamefont {C.}~\bibnamefont {Robert}}, \bibinfo {author} {\bibfnamefont {X.}~\bibnamefont {Marie}}, \bibinfo {author} {\bibfnamefont {B.}~\bibnamefont {Urbaszek}}, \bibinfo {author} {\bibfnamefont {A.}~\bibnamefont {Knorr}}, \ and\ \bibinfo {author} {\bibfnamefont {E.}~\bibnamefont {Malic}},\ }\bibfield  {title} {\enquote {\bibinfo {title} {Phonon-assisted photoluminescence from indirect excitons in monolayers of transition-metal dichalcogenides},}\ }\href {\doibase 10.1021/acs.nanolett.0c00633} {\bibfield  {journal} {\bibinfo  {journal} {Nano Letters}\ }\textbf {\bibinfo {volume} {20}},\ \bibinfo {pages} {2849--2856} (\bibinfo {year} {2020})}\BibitemShut
  {NoStop}%
\bibitem [{\citenamefont {Li}\ \emph {et~al.}(2020)\citenamefont {Li}, \citenamefont {Wang}, \citenamefont {Miao}, \citenamefont {Lian},\ and\ \citenamefont {Shi}}]{Li2020}%
  \BibitemOpen
  \bibfield  {author} {\bibinfo {author} {\bibfnamefont {Z.}~\bibnamefont {Li}}, \bibinfo {author} {\bibfnamefont {T.}~\bibnamefont {Wang}}, \bibinfo {author} {\bibfnamefont {S.}~\bibnamefont {Miao}}, \bibinfo {author} {\bibfnamefont {Z.}~\bibnamefont {Lian}}, \ and\ \bibinfo {author} {\bibfnamefont {S.-F.}\ \bibnamefont {Shi}},\ }\bibfield  {title} {\enquote {\bibinfo {title} {Fine structures of valley-polarized excitonic states in monolayer transitional metal dichalcogenides},}\ }\href {\doibase doi:10.1515/nanoph-2020-0054} {\bibfield  {journal} {\bibinfo  {journal} {Nanophotonics}\ }\textbf {\bibinfo {volume} {9}},\ \bibinfo {pages} {1811--1829} (\bibinfo {year} {2020})}\BibitemShut {NoStop}%
\bibitem [{\citenamefont {Ruppert}, \citenamefont {Aslan},\ and\ \citenamefont {Heinz}(2014)}]{Ruppert2014}%
  \BibitemOpen
  \bibfield  {author} {\bibinfo {author} {\bibfnamefont {C.}~\bibnamefont {Ruppert}}, \bibinfo {author} {\bibfnamefont {O.~B.}\ \bibnamefont {Aslan}}, \ and\ \bibinfo {author} {\bibfnamefont {T.~F.}\ \bibnamefont {Heinz}},\ }\bibfield  {title} {\enquote {\bibinfo {title} {Optical properties and band gap of single- and few-layer mote2 crystals},}\ }\href {\doibase 10.1021/nl502557g} {\bibfield  {journal} {\bibinfo  {journal} {Nano Letters}\ }\textbf {\bibinfo {volume} {14}},\ \bibinfo {pages} {6231--6236} (\bibinfo {year} {2014})},\ \bibinfo {note} {pMID: 25302768},\ \Eprint {http://arxiv.org/abs/https://doi.org/10.1021/nl502557g} {https://doi.org/10.1021/nl502557g} \BibitemShut {NoStop}%
\bibitem [{\citenamefont {Nozik}(2001)}]{Nozik2001}%
  \BibitemOpen
  \bibfield  {author} {\bibinfo {author} {\bibfnamefont {A.}~\bibnamefont {Nozik}},\ }\bibfield  {title} {\enquote {\bibinfo {title} {Nozik, a.j.: Spectroscopy and hot electron relaxation dynamics in semiconductor quantum wells and quantum dots. annu. rev. phys. chem. 52(1), 193-231},}\ }\href {\doibase 10.1146/annurev.physchem.52.1.193} {\bibfield  {journal} {\bibinfo  {journal} {Annual review of physical chemistry}\ }\textbf {\bibinfo {volume} {52}},\ \bibinfo {pages} {193--231} (\bibinfo {year} {2001})}\BibitemShut {NoStop}%
\bibitem [{\citenamefont {Li}\ \emph {et~al.}(2016)\citenamefont {Li}, \citenamefont {Wang}, \citenamefont {Gong}, \citenamefont {Yun}, \citenamefont {Zhang},\ and\ \citenamefont {Ding}}]{LI20161809}%
  \BibitemOpen
  \bibfield  {author} {\bibinfo {author} {\bibfnamefont {Q.}~\bibnamefont {Li}}, \bibinfo {author} {\bibfnamefont {S.}~\bibnamefont {Wang}}, \bibinfo {author} {\bibfnamefont {Z.-N.}\ \bibnamefont {Gong}}, \bibinfo {author} {\bibfnamefont {F.}~\bibnamefont {Yun}}, \bibinfo {author} {\bibfnamefont {Y.}~\bibnamefont {Zhang}}, \ and\ \bibinfo {author} {\bibfnamefont {W.}~\bibnamefont {Ding}},\ }\bibfield  {title} {\enquote {\bibinfo {title} {Time-resolved photoluminescence studies of ingan/gan multi-quantum-wells blue and green light-emitting diodes at room temperature},}\ }\href {\doibase https://doi.org/10.1016/j.ijleo.2015.11.095} {\bibfield  {journal} {\bibinfo  {journal} {Optik}\ }\textbf {\bibinfo {volume} {127}},\ \bibinfo {pages} {1809--1813} (\bibinfo {year} {2016})}\BibitemShut {NoStop}%
\bibitem [{\citenamefont {Syperek}\ \emph {et~al.}(2010)\citenamefont {Syperek}, \citenamefont {Leszczyński}, \citenamefont {Misiewicz}, \citenamefont {Pavelescu}, \citenamefont {Gilfert},\ and\ \citenamefont {Reithmaier}}]{doi:10.1063/1.3280384}%
  \BibitemOpen
  \bibfield  {author} {\bibinfo {author} {\bibfnamefont {M.}~\bibnamefont {Syperek}}, \bibinfo {author} {\bibfnamefont {P.}~\bibnamefont {Leszczyński}}, \bibinfo {author} {\bibfnamefont {J.}~\bibnamefont {Misiewicz}}, \bibinfo {author} {\bibfnamefont {E.~M.}\ \bibnamefont {Pavelescu}}, \bibinfo {author} {\bibfnamefont {C.}~\bibnamefont {Gilfert}}, \ and\ \bibinfo {author} {\bibfnamefont {J.~P.}\ \bibnamefont {Reithmaier}},\ }\bibfield  {title} {\enquote {\bibinfo {title} {Time-resolved photoluminescence spectroscopy of an ingaas/gaas quantum well-quantum dots tunnel injection structure},}\ }\href {\doibase 10.1063/1.3280384} {\bibfield  {journal} {\bibinfo  {journal} {Applied Physics Letters}\ }\textbf {\bibinfo {volume} {96}},\ \bibinfo {pages} {011901} (\bibinfo {year} {2010})},\ \Eprint {http://arxiv.org/abs/https://doi.org/10.1063/1.3280384} {https://doi.org/10.1063/1.3280384} \BibitemShut {NoStop}%
\bibitem [{\citenamefont {Okamoto}\ \emph {et~al.}(2005)\citenamefont {Okamoto}, \citenamefont {Niki}, \citenamefont {Scherer}, \citenamefont {Narukawa}, \citenamefont {Mukai},\ and\ \citenamefont {Kawakami}}]{Okamoto2005}%
  \BibitemOpen
  \bibfield  {author} {\bibinfo {author} {\bibfnamefont {K.}~\bibnamefont {Okamoto}}, \bibinfo {author} {\bibfnamefont {I.}~\bibnamefont {Niki}}, \bibinfo {author} {\bibfnamefont {A.}~\bibnamefont {Scherer}}, \bibinfo {author} {\bibfnamefont {Y.}~\bibnamefont {Narukawa}}, \bibinfo {author} {\bibfnamefont {T.}~\bibnamefont {Mukai}}, \ and\ \bibinfo {author} {\bibfnamefont {Y.}~\bibnamefont {Kawakami}},\ }\bibfield  {title} {\enquote {\bibinfo {title} {Surface plasmon enhanced spontaneous emission rate of ingan-gan quantum wells probed by time-resolved photoluminescence spectroscopy},}\ }\href {\doibase 10.1063/1.2010602} {\bibfield  {journal} {\bibinfo  {journal} {Applied Physics Letters}\ }\textbf {\bibinfo {volume} {87}},\ \bibinfo {pages} {071102} (\bibinfo {year} {2005})},\ \Eprint {http://arxiv.org/abs/https://doi.org/10.1063/1.2010602} {https://doi.org/10.1063/1.2010602} \BibitemShut {NoStop}%
\bibitem [{\citenamefont {Lefebvre}\ \emph {et~al.}(1999)\citenamefont {Lefebvre}, \citenamefont {Allègre}, \citenamefont {Gil}, \citenamefont {Mathieu}, \citenamefont {Bigenwald}, \citenamefont {Grandjean}, \citenamefont {Leroux},\ and\ \citenamefont {Massies}}]{Lefebvre1999}%
  \BibitemOpen
  \bibfield  {author} {\bibinfo {author} {\bibfnamefont {P.}~\bibnamefont {Lefebvre}}, \bibinfo {author} {\bibfnamefont {J.}~\bibnamefont {Allègre}}, \bibinfo {author} {\bibfnamefont {B.}~\bibnamefont {Gil}}, \bibinfo {author} {\bibfnamefont {H.}~\bibnamefont {Mathieu}}, \bibinfo {author} {\bibfnamefont {P.}~\bibnamefont {Bigenwald}}, \bibinfo {author} {\bibfnamefont {N.}~\bibnamefont {Grandjean}}, \bibinfo {author} {\bibfnamefont {M.}~\bibnamefont {Leroux}}, \ and\ \bibinfo {author} {\bibfnamefont {J.}~\bibnamefont {Massies}},\ }\bibfield  {title} {\enquote {\bibinfo {title} {Time-resolved photoluminescence as a probe of internal electric fields in gan-(gaa1)n quantum wells},}\ }\href {\doibase 10.1103/PhysRevB.59.15363} {\bibfield  {journal} {\bibinfo  {journal} {Physical review. B, Condensed matter}\ }\textbf {\bibinfo {volume} {59}},\ \bibinfo {pages} {15363} (\bibinfo {year} {1999})}\BibitemShut {NoStop}%
\bibitem [{\citenamefont {Modesti}\ \emph {et~al.}(1980)\citenamefont {Modesti}, \citenamefont {Frova}, \citenamefont {Capizzi}, \citenamefont {Quagliano},\ and\ \citenamefont {Staehli}}]{Modesti1980}%
  \BibitemOpen
  \bibfield  {author} {\bibinfo {author} {\bibfnamefont {S.}~\bibnamefont {Modesti}}, \bibinfo {author} {\bibfnamefont {A.}~\bibnamefont {Frova}}, \bibinfo {author} {\bibfnamefont {M.}~\bibnamefont {Capizzi}}, \bibinfo {author} {\bibfnamefont {L.~G.}\ \bibnamefont {Quagliano}}, \ and\ \bibinfo {author} {\bibfnamefont {J.-L.}\ \bibnamefont {Staehli}},\ }\bibfield  {title} {\enquote {\bibinfo {title} {E-h plasma luminescence in gaas sub(1-x)p sub(x) above direct-indirect crossover},}\ }\href {http://inis.iaea.org/search/search.aspx?orig_q=RN:13666555} {\bibfield  {journal} {\bibinfo  {journal} {Journal of the Physical Society of Japan}\ }\textbf {\bibinfo {volume} {49}},\ \bibinfo {pages} {515--518} (\bibinfo {year} {1980})}\BibitemShut {NoStop}%
\bibitem [{\citenamefont {Modesti}\ \emph {et~al.}(1981{\natexlab{a}})\citenamefont {Modesti}, \citenamefont {Quagliano}, \citenamefont {Frova}, \citenamefont {Staehli},\ and\ \citenamefont {Guzzi}}]{MODESTI1981581}%
  \BibitemOpen
  \bibfield  {author} {\bibinfo {author} {\bibfnamefont {S.}~\bibnamefont {Modesti}}, \bibinfo {author} {\bibfnamefont {L.}~\bibnamefont {Quagliano}}, \bibinfo {author} {\bibfnamefont {A.}~\bibnamefont {Frova}}, \bibinfo {author} {\bibfnamefont {J.-L.}\ \bibnamefont {Staehli}}, \ and\ \bibinfo {author} {\bibfnamefont {M.}~\bibnamefont {Guzzi}},\ }\bibfield  {title} {\enquote {\bibinfo {title} {High-excitation luminescence in direct-gap gaas1-xpx: E-h plasma expansion effects},}\ }\href {\doibase https://doi.org/10.1016/0022-2313(81)90045-4} {\bibfield  {journal} {\bibinfo  {journal} {Journal of Luminescence}\ }\textbf {\bibinfo {volume} {24-25}},\ \bibinfo {pages} {581--584} (\bibinfo {year} {1981}{\natexlab{a}})}\BibitemShut {NoStop}%
\bibitem [{\citenamefont {Modesti}\ \emph {et~al.}(1981{\natexlab{b}})\citenamefont {Modesti}, \citenamefont {Frova}, \citenamefont {Staehli}, \citenamefont {Guzzi},\ and\ \citenamefont {Capizzi}}]{Modesti1981exp2}%
  \BibitemOpen
  \bibfield  {author} {\bibinfo {author} {\bibfnamefont {S.}~\bibnamefont {Modesti}}, \bibinfo {author} {\bibfnamefont {A.}~\bibnamefont {Frova}}, \bibinfo {author} {\bibfnamefont {J.}~\bibnamefont {Staehli}}, \bibinfo {author} {\bibfnamefont {M.}~\bibnamefont {Guzzi}}, \ and\ \bibinfo {author} {\bibfnamefont {M.}~\bibnamefont {Capizzi}},\ }\bibfield  {title} {\enquote {\bibinfo {title} {Electron-hole plasma expansion in direct-gap gaas1-xpx},}\ }\href {\doibase 10.1002/pssb.2221080134} {\bibfield  {journal} {\bibinfo  {journal} {Physica Status Solidi B-basic Solid State Physics - PHYS STATUS SOLIDI B-BASIC SO}\ }\textbf {\bibinfo {volume} {108}},\ \bibinfo {pages} {281--288} (\bibinfo {year} {1981}{\natexlab{b}})}\BibitemShut {NoStop}%
\bibitem [{\citenamefont {Perri}\ \emph {et~al.}(2017)\citenamefont {Perri}, \citenamefont {Preda}, \citenamefont {D'Andrea}, \citenamefont {Thyrhaug}, \citenamefont {Cerullo}, \citenamefont {Polli},\ and\ \citenamefont {Hauer}}]{Perri2017}%
  \BibitemOpen
  \bibfield  {author} {\bibinfo {author} {\bibfnamefont {A.}~\bibnamefont {Perri}}, \bibinfo {author} {\bibfnamefont {F.}~\bibnamefont {Preda}}, \bibinfo {author} {\bibfnamefont {C.}~\bibnamefont {D'Andrea}}, \bibinfo {author} {\bibfnamefont {E.}~\bibnamefont {Thyrhaug}}, \bibinfo {author} {\bibfnamefont {G.}~\bibnamefont {Cerullo}}, \bibinfo {author} {\bibfnamefont {D.}~\bibnamefont {Polli}}, \ and\ \bibinfo {author} {\bibfnamefont {J.}~\bibnamefont {Hauer}},\ }\bibfield  {title} {\enquote {\bibinfo {title} {Excitation-emission fourier-transform spectroscopy based on a birefringent interferometer},}\ }\href {\doibase 10.1364/OE.25.00A483} {\bibfield  {journal} {\bibinfo  {journal} {Opt. Express}\ }\textbf {\bibinfo {volume} {25}},\ \bibinfo {pages} {A483--A490} (\bibinfo {year} {2017})}\BibitemShut {NoStop}%
\bibitem [{\citenamefont {Perri}\ \emph {et~al.}(2018)\citenamefont {Perri}, \citenamefont {Gaida}, \citenamefont {Farina}, \citenamefont {Preda}, \citenamefont {Viola}, \citenamefont {Ballottari}, \citenamefont {Hauer}, \citenamefont {Silvestri}, \citenamefont {D'Andrea}, \citenamefont {Cerullo},\ and\ \citenamefont {Polli}}]{Perri2018}%
  \BibitemOpen
  \bibfield  {author} {\bibinfo {author} {\bibfnamefont {A.}~\bibnamefont {Perri}}, \bibinfo {author} {\bibfnamefont {J.~H.}\ \bibnamefont {Gaida}}, \bibinfo {author} {\bibfnamefont {A.}~\bibnamefont {Farina}}, \bibinfo {author} {\bibfnamefont {F.}~\bibnamefont {Preda}}, \bibinfo {author} {\bibfnamefont {D.}~\bibnamefont {Viola}}, \bibinfo {author} {\bibfnamefont {M.}~\bibnamefont {Ballottari}}, \bibinfo {author} {\bibfnamefont {J.}~\bibnamefont {Hauer}}, \bibinfo {author} {\bibfnamefont {S.~D.}\ \bibnamefont {Silvestri}}, \bibinfo {author} {\bibfnamefont {C.}~\bibnamefont {D'Andrea}}, \bibinfo {author} {\bibfnamefont {G.}~\bibnamefont {Cerullo}}, \ and\ \bibinfo {author} {\bibfnamefont {D.}~\bibnamefont {Polli}},\ }\bibfield  {title} {\enquote {\bibinfo {title} {Time- and frequency-resolved fluorescence with a single tcspc detector via a fourier-transform approach},}\ }\href {\doibase 10.1364/OE.26.002270} {\bibfield  {journal} {\bibinfo  {journal} {Opt. Express}\ }\textbf {\bibinfo {volume} {26}},\ \bibinfo
  {pages} {2270--2279} (\bibinfo {year} {2018})}\BibitemShut {NoStop}%
\bibitem [{Ref(2022)}]{RefractiveIndex}%
  \BibitemOpen
  \href {https://refractiveindex.info/?shelf=main&book=GaAs&page=Aspnes} {\enquote {\bibinfo {title} {{RefractiveIndex.info}, https://refractiveindex.info/},}\ } (\bibinfo {year} {2022})\BibitemShut {NoStop}%
\bibitem [{\citenamefont {Deveaud}\ \emph {et~al.}(1988)\citenamefont {Deveaud}, \citenamefont {Shah}, \citenamefont {Damen}, \citenamefont {Gossard},\ and\ \citenamefont {Lugli}}]{DEVEAUD1988435}%
  \BibitemOpen
  \bibfield  {author} {\bibinfo {author} {\bibfnamefont {B.}~\bibnamefont {Deveaud}}, \bibinfo {author} {\bibfnamefont {J.}~\bibnamefont {Shah}}, \bibinfo {author} {\bibfnamefont {T.}~\bibnamefont {Damen}}, \bibinfo {author} {\bibfnamefont {A.}~\bibnamefont {Gossard}}, \ and\ \bibinfo {author} {\bibfnamefont {P.}~\bibnamefont {Lugli}},\ }\bibfield  {title} {\enquote {\bibinfo {title} {Initial relaxation of photoexcited carriers in gaas and gaas quantum wells under subpicosecond excitation},}\ }\href {\doibase https://doi.org/10.1016/0038-1101(88)90312-7} {\bibfield  {journal} {\bibinfo  {journal} {Solid-State Electronics}\ }\textbf {\bibinfo {volume} {31}},\ \bibinfo {pages} {435--438} (\bibinfo {year} {1988})}\BibitemShut {NoStop}%
\bibitem [{\citenamefont {{Mosko}}\ and\ \citenamefont {{K{\'a}lna}}(1999)}]{1999SeScT..14..790M}%
  \BibitemOpen
  \bibfield  {author} {\bibinfo {author} {\bibfnamefont {M.}~\bibnamefont {{Mosko}}}\ and\ \bibinfo {author} {\bibfnamefont {K.}~\bibnamefont {{K{\'a}lna}}},\ }\bibfield  {title} {\enquote {\bibinfo {title} {{Carrier capture into a GaAs quantum well with a separate confinement region: comment on quantum and classical aspects}},}\ }\href {\doibase 10.1088/0268-1242/14/9/308} {\bibfield  {journal} {\bibinfo  {journal} {Semiconductor Science Technology}\ }\textbf {\bibinfo {volume} {14}},\ \bibinfo {pages} {790--796} (\bibinfo {year} {1999})}\BibitemShut {NoStop}%
\bibitem [{IOF(2022)}]{IOFFRE}%
  \BibitemOpen
  \href {http://www.ioffe.ru/SVA/NSM/Semicond/AlGaAs/index.html} {\enquote {\bibinfo {title} {{Ioffre Institute}, http://www.ioffe.ru},}\ } (\bibinfo {year} {2022})\BibitemShut {NoStop}%
\bibitem [{\citenamefont {Uchiki}\ \emph {et~al.}(1985)\citenamefont {Uchiki}, \citenamefont {Arakawa}, \citenamefont {Sakaki},\ and\ \citenamefont {Kobayashi}}]{UCHIKI1985311}%
  \BibitemOpen
  \bibfield  {author} {\bibinfo {author} {\bibfnamefont {H.}~\bibnamefont {Uchiki}}, \bibinfo {author} {\bibfnamefont {Y.}~\bibnamefont {Arakawa}}, \bibinfo {author} {\bibfnamefont {H.}~\bibnamefont {Sakaki}}, \ and\ \bibinfo {author} {\bibfnamefont {T.}~\bibnamefont {Kobayashi}},\ }\bibfield  {title} {\enquote {\bibinfo {title} {Hot photoluminescence of gaas-algaas multiple quantum well structures under high excitation by a single shot of 30 ps, 532 nm laser},}\ }\href {\doibase https://doi.org/10.1016/0038-1098(85)90615-5} {\bibfield  {journal} {\bibinfo  {journal} {Solid State Communications}\ }\textbf {\bibinfo {volume} {55}},\ \bibinfo {pages} {311--315} (\bibinfo {year} {1985})}\BibitemShut {NoStop}%
\bibitem [{\citenamefont {Tr\"ankle}\ \emph {et~al.}(1987{\natexlab{a}})\citenamefont {Tr\"ankle}, \citenamefont {Leier}, \citenamefont {Forchel}, \citenamefont {Haug}, \citenamefont {Ell},\ and\ \citenamefont {Weimann}}]{PhysRevLett.58.419}%
  \BibitemOpen
  \bibfield  {author} {\bibinfo {author} {\bibfnamefont {G.}~\bibnamefont {Tr\"ankle}}, \bibinfo {author} {\bibfnamefont {H.}~\bibnamefont {Leier}}, \bibinfo {author} {\bibfnamefont {A.}~\bibnamefont {Forchel}}, \bibinfo {author} {\bibfnamefont {H.}~\bibnamefont {Haug}}, \bibinfo {author} {\bibfnamefont {C.}~\bibnamefont {Ell}}, \ and\ \bibinfo {author} {\bibfnamefont {G.}~\bibnamefont {Weimann}},\ }\bibfield  {title} {\enquote {\bibinfo {title} {Dimensionality dependence of the band-gap renormalization in two- and three-dimensional electron-hole plasmas in gaas},}\ }\href {\doibase 10.1103/PhysRevLett.58.419} {\bibfield  {journal} {\bibinfo  {journal} {Phys. Rev. Lett.}\ }\textbf {\bibinfo {volume} {58}},\ \bibinfo {pages} {419--422} (\bibinfo {year} {1987}{\natexlab{a}})}\BibitemShut {NoStop}%
\bibitem [{\citenamefont {Tr\"ankle}\ \emph {et~al.}(1987{\natexlab{b}})\citenamefont {Tr\"ankle}, \citenamefont {Leier}, \citenamefont {Forchel}, \citenamefont {Haug}, \citenamefont {Ell},\ and\ \citenamefont {Weimann}}]{Trankle1987-gs}%
  \BibitemOpen
  \bibfield  {author} {\bibinfo {author} {\bibfnamefont {G.}~\bibnamefont {Tr\"ankle}}, \bibinfo {author} {\bibfnamefont {H.}~\bibnamefont {Leier}}, \bibinfo {author} {\bibfnamefont {A.}~\bibnamefont {Forchel}}, \bibinfo {author} {\bibfnamefont {H.}~\bibnamefont {Haug}}, \bibinfo {author} {\bibfnamefont {C.}~\bibnamefont {Ell}}, \ and\ \bibinfo {author} {\bibfnamefont {G.}~\bibnamefont {Weimann}},\ }\bibfield  {title} {\enquote {\bibinfo {title} {Dimensionality dependence of the band-gap renormalization in two- and three-dimensional electron-hole plasmas in gaas},}\ }\href {\doibase 10.1103/PhysRevLett.58.419} {\bibfield  {journal} {\bibinfo  {journal} {Phys. Rev. Lett.}\ }\textbf {\bibinfo {volume} {58}},\ \bibinfo {pages} {419--422} (\bibinfo {year} {1987}{\natexlab{b}})}\BibitemShut {NoStop}%
\bibitem [{\citenamefont {Di~Cicco}\ \emph {et~al.}(2020)\citenamefont {Di~Cicco}, \citenamefont {Polzoni}, \citenamefont {Gunnella}, \citenamefont {Trapananti}, \citenamefont {Minicucci}, \citenamefont {Rezvani}, \citenamefont {Catone}, \citenamefont {Di~Mario}, \citenamefont {Pelli~Cresi}, \citenamefont {Turchini},\ and\ \citenamefont {Martelli}}]{DiCicco2020}%
  \BibitemOpen
  \bibfield  {author} {\bibinfo {author} {\bibfnamefont {A.}~\bibnamefont {Di~Cicco}}, \bibinfo {author} {\bibfnamefont {G.}~\bibnamefont {Polzoni}}, \bibinfo {author} {\bibfnamefont {R.}~\bibnamefont {Gunnella}}, \bibinfo {author} {\bibfnamefont {A.}~\bibnamefont {Trapananti}}, \bibinfo {author} {\bibfnamefont {M.}~\bibnamefont {Minicucci}}, \bibinfo {author} {\bibfnamefont {S.~J.}\ \bibnamefont {Rezvani}}, \bibinfo {author} {\bibfnamefont {D.}~\bibnamefont {Catone}}, \bibinfo {author} {\bibfnamefont {L.}~\bibnamefont {Di~Mario}}, \bibinfo {author} {\bibfnamefont {J.~S.}\ \bibnamefont {Pelli~Cresi}}, \bibinfo {author} {\bibfnamefont {S.}~\bibnamefont {Turchini}}, \ and\ \bibinfo {author} {\bibfnamefont {F.}~\bibnamefont {Martelli}},\ }\bibfield  {title} {\enquote {\bibinfo {title} {Broadband optical ultrafast reflectivity of si, ge and gaas},}\ }\href {\doibase 10.1038/s41598-020-74068-y} {\bibfield  {journal} {\bibinfo  {journal} {Scientific Reports}\ }\textbf {\bibinfo {volume} {10}},\ \bibinfo {pages}
  {17363} (\bibinfo {year} {2020})}\BibitemShut {NoStop}%
\bibitem [{\citenamefont {Xing}\ \emph {et~al.}(2017)\citenamefont {Xing}, \citenamefont {Wang}, \citenamefont {Yang}, \citenamefont {Wang}, \citenamefont {Hao}, \citenamefont {Sun}, \citenamefont {Xiong}, \citenamefont {Luo}, \citenamefont {Han}, \citenamefont {Wang},\ and\ \citenamefont {Li}}]{Xing2017}%
  \BibitemOpen
  \bibfield  {author} {\bibinfo {author} {\bibfnamefont {Y.}~\bibnamefont {Xing}}, \bibinfo {author} {\bibfnamefont {L.}~\bibnamefont {Wang}}, \bibinfo {author} {\bibfnamefont {D.}~\bibnamefont {Yang}}, \bibinfo {author} {\bibfnamefont {Z.}~\bibnamefont {Wang}}, \bibinfo {author} {\bibfnamefont {Z.}~\bibnamefont {Hao}}, \bibinfo {author} {\bibfnamefont {C.}~\bibnamefont {Sun}}, \bibinfo {author} {\bibfnamefont {B.}~\bibnamefont {Xiong}}, \bibinfo {author} {\bibfnamefont {Y.}~\bibnamefont {Luo}}, \bibinfo {author} {\bibfnamefont {Y.}~\bibnamefont {Han}}, \bibinfo {author} {\bibfnamefont {J.}~\bibnamefont {Wang}}, \ and\ \bibinfo {author} {\bibfnamefont {H.}~\bibnamefont {Li}},\ }\bibfield  {title} {\enquote {\bibinfo {title} {A novel model on time-resolved photoluminescence measurements of polar ingan/gan multi-quantum-well structures},}\ }\href {\doibase 10.1038/srep45082} {\bibfield  {journal} {\bibinfo  {journal} {Scientific Reports}\ }\textbf {\bibinfo {volume} {7}},\ \bibinfo {pages} {45082} (\bibinfo
  {year} {2017})}\BibitemShut {NoStop}%
\bibitem [{\citenamefont {Ilahi}(2023)}]{ILAHI2023414612}%
  \BibitemOpen
  \bibfield  {author} {\bibinfo {author} {\bibfnamefont {S.}~\bibnamefont {Ilahi}},\ }\bibfield  {title} {\enquote {\bibinfo {title} {Doping effects on minority carrier parameters in bulk gaas},}\ }\href {\doibase https://doi.org/10.1016/j.physb.2022.414612} {\bibfield  {journal} {\bibinfo  {journal} {Physica B: Condensed Matter}\ }\textbf {\bibinfo {volume} {652}},\ \bibinfo {pages} {414612} (\bibinfo {year} {2023})}\BibitemShut {NoStop}%
\bibitem [{\citenamefont {Fouquet}\ and\ \citenamefont {Burnham}(1986)}]{1073154}%
  \BibitemOpen
  \bibfield  {author} {\bibinfo {author} {\bibfnamefont {J.}~\bibnamefont {Fouquet}}\ and\ \bibinfo {author} {\bibfnamefont {R.}~\bibnamefont {Burnham}},\ }\bibfield  {title} {\enquote {\bibinfo {title} {Recombination dynamics in gaas/alxga1- xas quantum well structures},}\ }\href {\doibase 10.1109/JQE.1986.1073154} {\bibfield  {journal} {\bibinfo  {journal} {IEEE Journal of Quantum Electronics}\ }\textbf {\bibinfo {volume} {22}},\ \bibinfo {pages} {1799--1810} (\bibinfo {year} {1986})}\BibitemShut {NoStop}%
\bibitem [{\citenamefont {Zhang}\ \emph {et~al.}(2022)\citenamefont {Zhang}, \citenamefont {Castaneda}, \citenamefont {Gfroerer}, \citenamefont {Friedman}, \citenamefont {Zhang}, \citenamefont {Wanlass},\ and\ \citenamefont {Zhang}}]{Zhang2022}%
  \BibitemOpen
  \bibfield  {author} {\bibinfo {author} {\bibfnamefont {F.}~\bibnamefont {Zhang}}, \bibinfo {author} {\bibfnamefont {J.~F.}\ \bibnamefont {Castaneda}}, \bibinfo {author} {\bibfnamefont {T.~H.}\ \bibnamefont {Gfroerer}}, \bibinfo {author} {\bibfnamefont {D.}~\bibnamefont {Friedman}}, \bibinfo {author} {\bibfnamefont {Y.-H.}\ \bibnamefont {Zhang}}, \bibinfo {author} {\bibfnamefont {M.~W.}\ \bibnamefont {Wanlass}}, \ and\ \bibinfo {author} {\bibfnamefont {Y.}~\bibnamefont {Zhang}},\ }\bibfield  {title} {\enquote {\bibinfo {title} {An all optical approach for comprehensive in-operando analysis of radiative and nonradiative recombination processes in gaas double heterostructures},}\ }\href {\doibase 10.1038/s41377-022-00833-5} {\bibfield  {journal} {\bibinfo  {journal} {Light: Science {\&} Applications}\ }\textbf {\bibinfo {volume} {11}},\ \bibinfo {pages} {137} (\bibinfo {year} {2022})}\BibitemShut {NoStop}%
\end{thebibliography}%

\end{document}